# Effect of mobile financial services on financial behavior in developing economies – Evidence from India

Shreya Biswas

Assistant Professor

Department of Economics and Finance

Birla Institute of Technology and Science, Pilani, Hyderabad Campus

India

E-mail: shreya@hyderabad.bits-pilani.ac.in

**Effect of mobile financial services on financial behavior in developing economies – Evidence from India**

## Abstract

*The study examines the relationship between mobile financial services and individual financial behavior in India wherein a sizeable population is yet to be financially included. Addressing the endogeneity associated with the use of mobile financial services using an instrumental variable method, the study finds that the use of mobile financial services increases the likelihood of investment, having insurance and borrowing from formal financial institutions. Further, the analysis highlights that access to mobile financial services have the potential to bridge the gender divide in financial inclusion. Fastening the pace of access to mobile financial services may partially alter pandemic induced poverty.*

1. Introduction

There exists a large gap in bank account ownership in developed and developing economies (Global Findex Report, 2018). Further, almost half of the world's unbanked population resides in Bangladesh, China, India, Mexico, Kenya and Nigeria. The extant literature indicates that financial inclusion[1] is related to an increase in welfare. At the macroeconomic level, financial inclusion fosters economic growth (Sahay et al., 2015), reduce poverty in developing economies (Burgess and Pande, 2005), reduce inequality (Demiriguc-Kunt and Levine, 2009; Neaime and Gaysset, 2018) as well as reduce carbon emissions (Renzhi and Baek, 2020). Among microeconomic outcomes, most striking are the positive effects of financial inclusion of women on health outcomes (Prina, 2015), women's position in the household (Ashraf et al., 2010) and better children's outcomes (Duflo, 2003).

Given empirical evidence on the multitude of positive effects of financial inclusion on economic outcomes, studies have examined what interventions can foster faster financial inclusion. Literature suggests that stronger legal rights, proximity to financial institutions and a stable political environment (Allen et al., 2016) is conducive for access and use of bank accounts, financial literacy is another channel that can improve financial inclusion (Grohman et al., 2018), social trust improves the use of basic financial services (Xu, 2020). In recent times, policymakers are of the view that digital financial services (DFS) and financial technology (fintech) has the potential to fasten the pace of financial inclusion in developing economies as it reduces the cost of transactions, improves trust, and increases the speed of transaction (World Bank, 2020). Mobile financial services increase the likelihood of savings and the amount saved by individuals in Africa (Ouma et al., 2017; Loaba, 2021). Fanta and Makina (2019) find that internet access and usage of mobile phones increase the usage of financial services like ATMs in African countries. In the context of India, Ghosh (2017) found that individuals with mobile phones are more likely to own bank accounts than non-mobile phone users. The evidence related to the use of mobile for financial transactions and its effect on financial outcomes is still scarce, especially outside Africa. This study intends to address this question in the context of a developing economy like India, which is home to a large proportion of the financially excluded population in the world. The question becomes even more important in the light of the current coronavirus pandemic as it threatens to push over 100 million people into poverty by the end of 2021, and financial inclusion has the potential to offset this pandemic-induced rise in poverty (Gutiérrez-Romero and Ahamed, 2021).

[1] Financial inclusion in its broadest sense refers to ability of individuals to access various financial service not limited to access to deposit account, formal credit and insurance.

The Pradhan Mantri Jan Dhan Yojana (PMJDY) introduced in 2014 is the flagship financial inclusion program in India with an objective to ensure each household can access low cost saving bank account along with financial literacy, access to formal credit, insurance and pension with over 424 million beneficiaries[2]. However, PMJDY followed a bank driven financial inclusion model. The Unified Payments Interface (UPI) launched in 2016 was one of the early policy interventions that pushed the growth of DFS in India. The fintech companies in India like PhonePe, Paytm, along with payment banks have led the DFS growth. This was further facilitated by reduced internet cost and the availability of low-cost smartphones in the last few years in the country. Given this transition, albeit a slow one from brick and mortar bank-based model to the growing importance of DFS in the country, taking stock of whether the use of mobile financial services improves financial behavior becomes an important question.

This study finds that the use of mobile financial services improves the likelihood of having insurance along with increasing the probability of investment and borrowing from formal institutions in India. Further, the relationship is largely similar across age groups and gender suggesting that the positive effect of mobile financial services may not be exclusionary. The findings of the paper underscore the importance of accelerating digital financial services to fasten the pace of financial inclusion in the country.

The rest of the paper is organized as follows. Section 2 presents the data and the variables of the study and section 3 discuss the methodology employed. Section 4 elaborates the results and section 5 summarizes the findings and concludes.

## 2. Data and variables

The study utilizes the data provided by the Financial Inclusion Insights (FII) program by Kantar, which is supported by the Bill and Melinda Gates Foundation. The FII program conducts a nationally representative survey in eight Asian and African countries. Our analysis is based on the most recent sixth wave of FII program in India. The survey was conducted between September to December 2018, covering 48,027 individuals aged 15 years and above[3].

[2] https://pmjdy.gov.in/home (accessed on June 15, 2021).
[3] https://finclusion.org/about/

The study considers three financial outcomes to reflect the investment, credit behavior and risk management practices of individuals. The first outcome is whether the individual is investing in financial products. The *Investment* variable is a binary outcome that takes the value one if the individual invested in either local shares, foreign shares, bonds, chit funds, land or gems, and jewelry. The second outcome variable is *Insurance* which takes the value one if the individual has life insurance and zero otherwise. The third outcome is related to the borrowing behavior of individuals. The *Borrowing* dummy takes the value one if the individual borrowed from formal financial institutions like any bank, post-office, using a card, self-help group or microfinance institutions and zero for borrowing from moneylenders or friends and relatives.

The interest variable is the use of mobile financial services (*MFS*) variable, a dummy that takes the value one if the individual used mobile for paying or receiving money and zero otherwise. We control for various socio-economic characteristics including age, level of education, having a bank account, marital status, gender, economic status of the household, religion, area of residence and state. Table 1 provides the definition of the variables used in the analysis.

[Insert Table 1 here]

Table 2 provides the summary statistics of the outcome variables, MFS and the other controls for the full sample (column 1) and separately for the MFS adopters (column 2) and MFS non-users (column 3). Less than 10% use mobile for financial transactions and the share of individuals who invest is also quite low. The insurance penetration is also low and among the borrowers, around 17% borrow from formal sources. The descriptive statistics indicate that there is significant scope to improve the financial behavior of individuals. Interestingly, we observe that the share of individuals with favorable outcomes is higher among the sample of MFS users compared to non-users. The univariate statistics suggest that MFS can possibly improve the financial outcomes in developing economies and needs further exploration.

[Insert Table 2 here]

## 3. Methodology

The financial outcomes considered in our analysis are binary, which warrants employing probit specification for analyzing the relationship between the MFS variable and various financial outcomes. However, in the presence of important omitted variables or reverse causality the estimator can be biased. One may argue that individuals who are aware are the MFS adopters as well as borrow from formal financial institutions or buy insurance. The study employs an instrumental variable approach to address the endogeneity. An instrument is a variable that should be correlated with the endogenous explanatory variable and uncorrelated with the financial outcome variables. The two instruments considered for the analysis are self-reported ability to adapt to technology and the share of higher education in the town/ village. The first instrument gives the ease with which an individual can browse the internet on mobile, download any application or use mobile for sending or receiving money. The three questions have four possible responses ranging from no ability to complete ability. In case the individual reports having some or complete ability in either of these three dimensions, then the technical ability variable takes the value one and zero otherwise. The self-reported ability of individuals to use mobile should be positively correlated with the likelihood of using mobile for financial transactions but should be otherwise unrelated to the financial outcomes. The second instrument is a continuous variable defined as the number of individuals in the town/ village who have completed at least higher secondary, excluding the individual herself/ himself. If in the vicinity, a higher share of individuals has a threshold level of education that may increase the pace of adoption of MFS but should be otherwise unrelated to an individual's financial outcomes.

The study adopts an instrumental variable two-stage least squares method (IV-2SLS) to examine the relationship using a linear probability model in both the first and second stages. The linear probability model will be less efficient than the probit specification; however, the linear approximation reports first stage F-statistic providing evidence regarding the strength of the instrument. The first stage linear probability model is given as:

$$MFS_i = \alpha_0 + \alpha_1 Tech\ ability_i + \alpha_2 Share\ of\ higher\ education_i + \sum \theta_k X_{ik} + State\ dummies + District\ dummies + u_i \qquad (1)$$

Where $Tech\ ability_i$ and $Share\ of\ higher\ education_i$ are the instruments that captures the individual's ability to adopt technology and the share of individuals in the town or village having at least completed higher secondary education respectively. The instrument validity condition requires that the parameter estimates of $\alpha_1$ and $\alpha_2$ are both positive and significant. $X_{ik}$ are the set of control variables that may affect financial outcomes. The state dummies capture side factors like state-specific policies that act as push factors for improving financial outcomes of individuals or the number of financial firms in the administrative area.

The second stage linear probability model is given by equation (2) below:

$$Y_i = \beta_0 + \beta_1 \widehat{MFS_i} + \sum \delta_k X_{ik} + State\ dummies + District\ dummies + e_i \quad (2)$$

Where $Y_i$ refers to the three financial outcomes and $\widehat{Fintech_i}$ is the predicted fincteh adoption value obtained after estimating equation (1). $\beta_1$ captures the average effect of MFS on financial outcomes.

Additionally, the paper also reports the instrumental variable probit model that uses a linear approximation in the first stage given by (1), but a probit specification in the second stage. The errors are clustered at the neighborhood level in all the specifications.

4. Results

4.1 Main results

Figure 1 gives the average marginal effect of MFS adoption on the three financial outcomes of individuals obtained from estimating probit models. It appears that MFS adoption at the individual level is positively associated with all the three financial outcomes at 1% level of significance. However, given the possibility that the MFS variable can be endogenous, we do not interpret the probit models further.

[Insert Figure 1 here]

Next, Table 3 reports the IV-2SLS output wherein the linear probability model is estimated in both first and second stages (equations 1 and 2). Specifically, we find that using MFS increases the probability of investment by 5.4 percentage points (column 1). The first stage regression output (column 2) indicates that the coefficients of tech ability and share of higher education are positive and significant at a 1% level of significance suggesting the instruments are positive predictors of the MFS variable. Further, the first stage F-stage value is much larger than the Stock-Yogo critical value reiterating that the instruments are not weak. The Hansen J-statistics over-identification test statistic is also insignificant at a 5% level of significance providing evidence regarding the validity of our instruments. Also, MFS improves insurance uptake and borrowing from formal financial institutions by 7.2% (column 3) and7.8%[4] (column 5), respectively.

[Table 3 here]

[4] The results of formal finance outcome need to be interpreted with caution as this pertains to only individuals who borrowed during the period and excludes non-borrowers from the analysis.

With respect to controls, we find that age of the individual, being married and having a bank account is positively related to the financial outcomes. Education at higher levels appears to have a positive relationship with financial outcomes. Further, female individuals are more likely to have life insurance, but less likely to borrow from formal financial institutions. Salaried individuals are more likely to invest and have insurance.

Table 4 reports the coefficients of IV-probit models, and the results are qualitatively similar to the results obtained using the linear approximation. The instruments are positively correlated to the mobile dummy at 1% level of significance, again reiterating the relevance of the instruments.

[Insert Table 4 here]

## 4.2 Heterogeneous effects

### 4.3.1 Female and male

There is a strong digital divide between males and females in India. Further, in Global South the decision to spend money is generally with the male earning member. Hence, it is possible that the main results can be driven by males in the sample. Figure 2 presents the marginal effect of MFS separately for females and males. There appears to be no difference across males and females in the relationship between MFS and investment and insurance variables. However, there is no discernable effect of MFS on formal borrowing behavior for males even though there is a positive relationship for females. This provides further evidence in favor of reducing the gender digital divide as it can improve the financial outcomes of females in the country.

[Insert Figure 2 here]

### 4.3.2 Younger and older age-cohorts

The younger individuals are likely to comprise a higher share of MFS adopters in the sample. However, younger individuals may have inferior financial outcomes given lower-income and more dependents at home. Sub-sample analysis of individuals below 45 years and older indicates that the positive effect of MFS is observed for both the age-cohorts even though the effect is more consistent for the younger cohort (Figure 3). There is no effect of MFS on the borrowing behavior of older adults in the sample.

[Insert Figure 3 here]

# 5 Discussion and conclusion

The study finds that the use of mobile financial services improves the financial outcomes of individuals in the form of a higher likelihood of investment, borrowing from formal financial institutions, and having insurance and this relationship appears to be consistent across gender and age groups.

The results highlight the need to fasten the pace of mobile financial services in developing countries of Asia and Africa as it has the potential of improving the financial behavior of individuals. Further, digital literacy programs especially targeting females, can be one of the means to reduce the gender gap in financial inclusion in India. Prioritizing access to mobile financial services, especially among the poor, may partially offset the rise in poverty due to COVID-19 lockdowns and health shock in an economy where a large proportion of the population is engaged in informal employment.

In spite of the policy implications, the study has a few limitations. In the absence of longitudinal data, the paper is unable to comment on whether overtime improvements in access to mobile financial services have a differential marginal effect on financial behavior. Secondly, accounting for other supply-side factors, including internet speed, cost, and access to electricity, remain another important dimension to identify the additional vulnerabilities and devise targeted policies for attaining the twin objectives of greater outreach of digital financial services and make financial inclusion pro-poor.

Figure 1: Marginal effects of mobile financial services obtained from probit

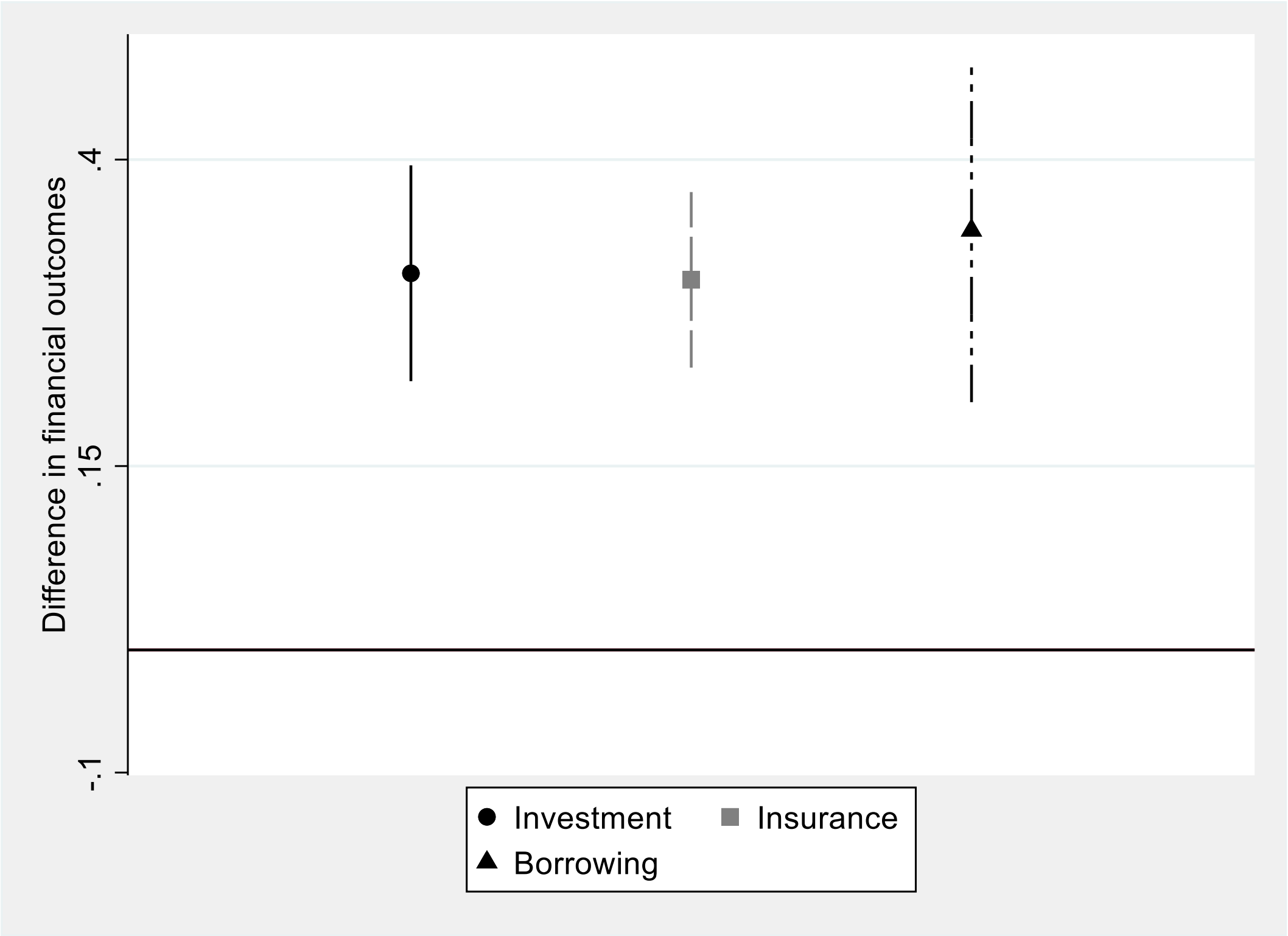

Figure 2: Gender-wise effect of mobile financial services on financial outcomes

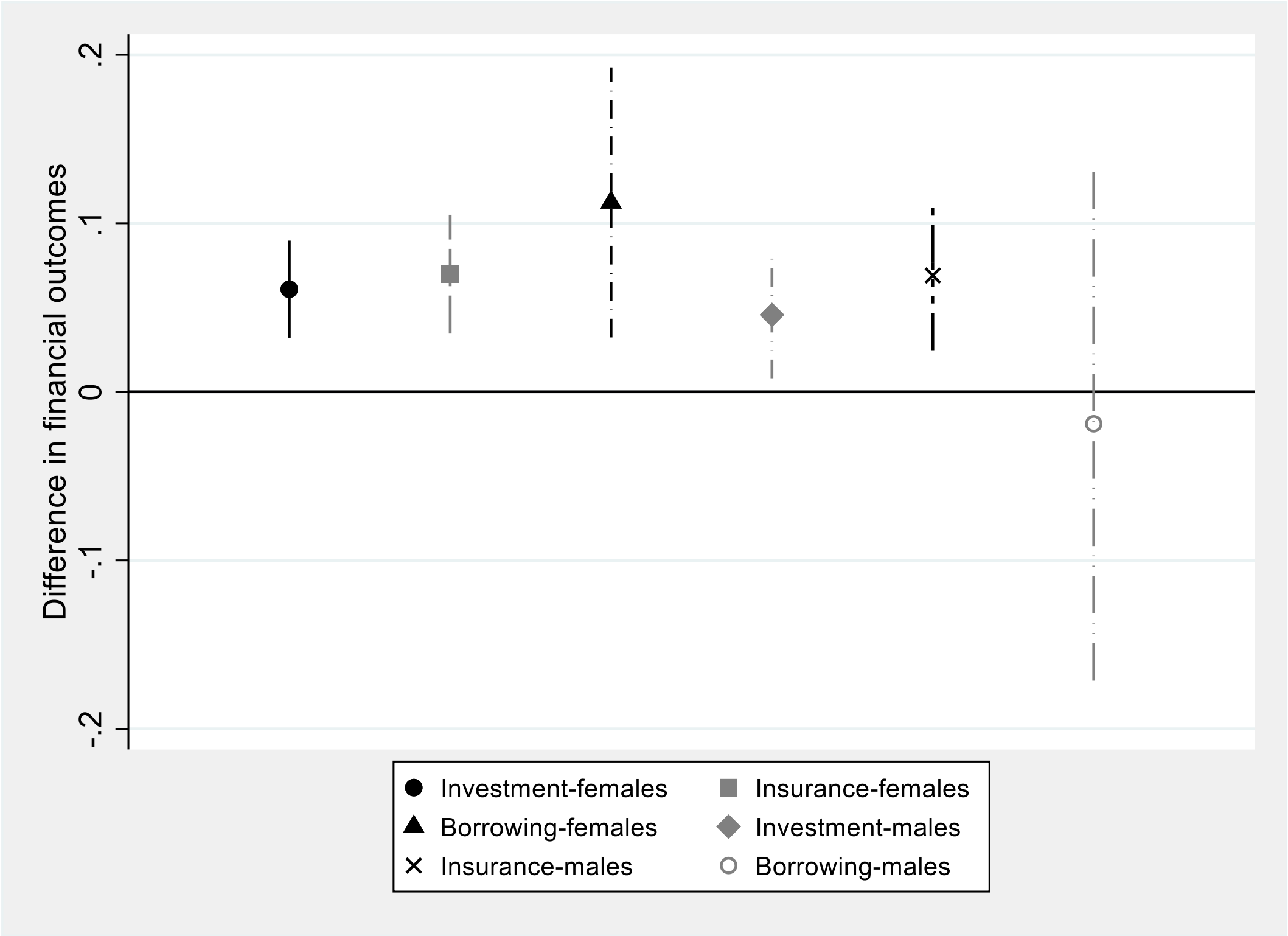

Figure 3: Age-wise effect of mobile financial services on financial outcomes

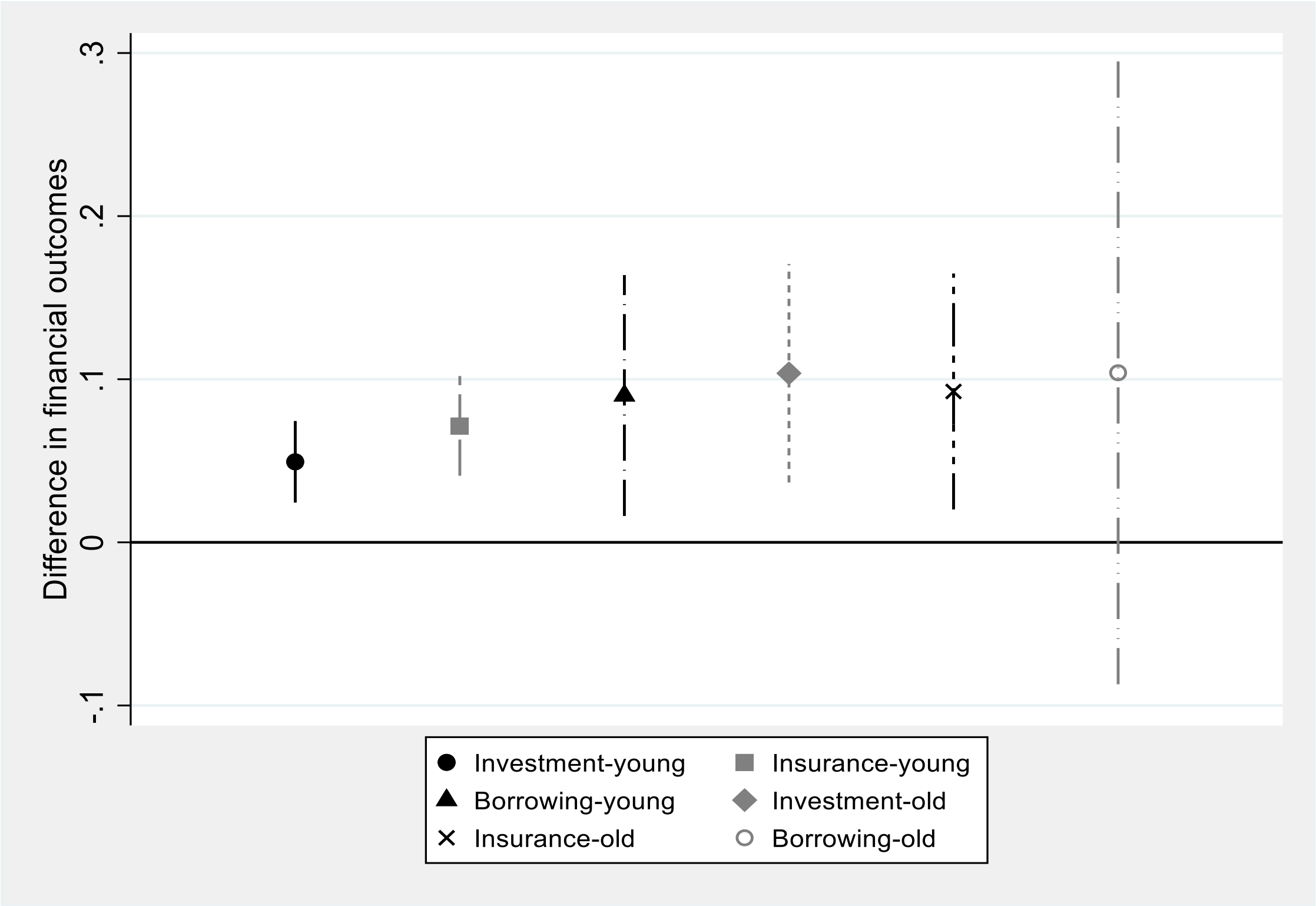

Table 1: Variable definitions

| Variable | Description |
|---|---|
| ***Dependent variable*** | |
| *Investment* | Dummy variable=1 if individual invests and zero for non-investors |
| *Insurance* | Dummy variable=1 if individual has a life insurance and zero for others |
| *Borrowing* | Dummy variable=1 if individual borrows from banks, non-banking financial institutions, using credit card, self-help groups and from microfinance institutions and zero for borrowing from friends, relatives and moneylenders |
| ***Interest variable*** | |
| *Mobile financial services (MFS)* | Dummy variable=1 if the individual uses mobile for financial transactions and zero for non-users |
| ***Controls*** | |
| *Age* | Age of the individual in years |
| *Married* | Marital status, 1 for married and zero for unmarried |
| *Urban* | Dummy variable=1 if the individual resides in urban areas and zero if the individual resides in rural area |
| *Bank account* | Dummy variable=1 if the individual has a bank account and zero otherwise |
| *Female* | Dummy variable=1 if the individual is a female and zero for males |
| *Non-poor* | Dummy variable=1 for non-poor and zero for poor |
| *Religion* | Religious identity of the individual- Hindu, Muslims or Others |
| *Education* | Educational attainment of individuals – Illiterate, literate with no formal education, <5years of schooling, primary completed (5 years of schooling), middle school (6-8 years), secondary completed (10 years), higher secondary completed (12 years), non-technical diploma, technical diploma, graduate or post graduate |
| *Occupation* | Occupational status of the individuals: Not working, Agriculture, Part-time, Non-farm business or salaried |

Table 2: Summary statistics

| | Full | MFS | No MFS |
|---|---|---|---|
| Dependent variable | | | |
| *Investment (%)* | 6.15 | 10.71 | 5.67 |
| *Insurance (%)* | 9.83 | 18.76 | 8.89 |
| *Borrowing (%)* | 17.86 | 25.95 | 17.07 |
| Interest variable | | | |
| *MFS (%)* | 9.58 | | |
| Controls | | | |
| *Age (in years)* | 37.69 | 31.82 | 38.32 |
| *Married (%)* | 80.95 | 65.32 | 82.61 |
| *Urban (%)* | 31.17 | 49.23 | 29.26 |
| *Bank account (%)* | 77.22 | 79.30 | 77.10 |
| *Female (%)* | 47.57 | 66.23 | 45.60 |
| *Non-poor (%)* | 68.15 | 41.66 | 70.95 |
| Religion (share in %) | | | |
| *Hindu* | 86.94 | 87.40 | 87.06 |
| *Muslim* | 10.07 | 7.67 | 10.35 |
| *Others* | 2.98 | 4.93 | 2.59 |
| Education categories (share in %) | | | |
| *No formal education* | 28.48 | 6.68 | 30.79 |
| *Literate – no formal education* | 4.04 | 2.88 | 4.17 |
| *<5years* | 6.26 | 4.19 | 6.48 |
| *Primary completed* | 8.14 | 5.35 | 8.44 |
| *Middle school* | 18.32 | 13.18 | 18.86 |
| *Secondary completed* | 15.62 | 17.31 | 15.44 |
| *Higher secondary completed* | 10.92 | 20.48 | 9.91 |
| *Non-technical diploma* | 1.33 | 5.02 | 0.94 |
| *Technical diploma* | 0.69 | 3.17 | 0.43 |
| *Graduate* | 4.95 | 16.57 | 3.72 |
| *Post graduate* | 1.24 | 5.17 | 0.82 |
| Occupation (share in %) | | | |
| *Not working* | 48.79 | 32.79 | 50.50 |
| *Agriculture* | 15.14 | 22.34 | 14.37 |
| *Irregular/ Part time work* | 9.18 | 7.27 | 9.34 |
| *Non-farm business* | 9.80 | 11.32 | 9.63 |
| *Salaried* | 17.12 | 26.28 | 16.15 |

Table 3: Effect of mobile financial services on financial outcomes – IV-2SLS output

| | (1) Investment- 2nd Stage | (2) Investment- 1st stage | (3) Insurance- 2nd Stage | (4) Insurance- 1st stage | (5) Borrowing- 2nd Stage | (6) Borrowing- 1st Stage |
|---|---|---|---|---|---|---|
| *MFS* | 0.054*** | | 0.072*** | | 0.078** | |
| | (0.013) | | (0.015) | | (0.037) | |
| *Tech ability* | | 0.387*** | | 0.387*** | | 0.439*** |
| | | (0.010) | | (0.010) | | (0.022) |
| *Share high education* | | 0.086*** | | 0.086*** | | 0.090*** |
| | | (0.022) | | (0.022) | | (0.034) |
| *Age* | 0.000*** | -0.000*** | 0.000** | -0.000*** | 0.001*** | -0.001*** |
| | (0.000) | (0.000) | (0.000) | (0.000) | (0.000) | (0.000) |
| *Bank account* | 0.022*** | -0.009** | 0.045*** | -0.009** | 0.080*** | 0.009 |
| | (0.003) | (0.004) | (0.003) | (0.004) | (0.009) | (0.007) |
| *Female* | 0.008*** | 0.007** | 0.013*** | 0.007** | -0.106*** | 0.004 |
| | (0.003) | (0.003) | (0.004) | (0.003) | (0.012) | (0.007) |
| *Non-poor* | -0.011*** | -0.028*** | -0.010** | -0.028*** | -0.040*** | -0.017** |
| | (0.004) | (0.004) | (0.004) | (0.004) | (0.011) | (0.007) |
| *Married* | 0.021*** | -0.019*** | 0.062*** | -0.019*** | 0.034** | -0.036*** |
| | (0.004) | (0.005) | (0.004) | (0.005) | (0.013) | (0.013) |
| *Education (Base: Illiterate)* | | | | | | |
| *Literate – no formal education* | 0.004 | -0.003 | 0.018** | -0.003 | 0.046** | 0.014 |
| | (0.007) | (0.006) | (0.008) | (0.006) | (0.022) | (0.012) |
| *<5years* | 0.002 | 0.004 | 0.012** | 0.004 | 0.005 | 0.015 |
| | (0.005) | (0.005) | (0.006) | (0.005) | (0.016) | (0.009) |
| *Primary completed* | 0.001 | 0.002 | 0.021*** | 0.002 | 0.030* | 0.005 |
| | (0.005) | (0.004) | (0.005) | (0.004) | (0.016) | (0.008) |
| *Middle school* | 0.003 | 0.002 | 0.034*** | 0.002 | 0.035*** | -0.002 |
| | (0.004) | (0.004) | (0.005) | (0.004) | (0.012) | (0.006) |
| *Secondary completed* | 0.007 | 0.003 | 0.059*** | 0.003 | 0.048*** | 0.010 |
| | (0.005) | (0.005) | (0.006) | (0.005) | (0.014) | (0.008) |
| *Higher* | 0.008 | 0.034*** | 0.075*** | 0.034*** | 0.040** | 0.034*** |
| *secondary completed* | | | | | | |
| | (0.005) | (0.006) | (0.007) | (0.006) | (0.017) | (0.012) |
| *Non-technical diploma* | 0.014 | 0.128*** | 0.060*** | 0.128*** | 0.079** | 0.128*** |
| | (0.012) | (0.018) | (0.015) | (0.018) | (0.037) | (0.032) |
| *Technical diploma* | -0.012 | 0.161*** | 0.088*** | 0.161*** | 0.079 | 0.281*** |
| | (0.015) | (0.025) | (0.023) | (0.025) | (0.064) | (0.057) |
| *Graduate* | 0.031*** | 0.109*** | 0.137*** | 0.109*** | 0.101*** | 0.158*** |
| | (0.008) | (0.009) | (0.011) | (0.009) | (0.027) | (0.023) |
| *Post graduate* | 0.039*** | 0.165*** | 0.162*** | 0.165*** | 0.254*** | 0.189*** |
| | (0.015) | (0.019) | (0.020) | (0.019) | (0.055) | (0.044) |
| *Occupation (Base: Not working)* | 0.004 | -0.003 | 0.018** | -0.003 | | 0.014 |
| *Agriculture* | 0.005 | 0.061*** | 0.012** | 0.061*** | 0.007 | 0.036*** |
| | (0.005) | (0.006) | (0.005) | (0.006) | (0.012) | (0.009) |
| *Part-time* | 0.031*** | 0.032*** | 0.028*** | 0.032*** | 0.077*** | 0.024*** |
| | (0.007) | (0.005) | (0.006) | (0.005) | (0.018) | (0.008) |
| *Non-farm business* | 0.007 | 0.029*** | 0.041*** | 0.029*** | 0.058*** | 0.018* |
| | (0.005) | (0.006) | (0.006) | (0.006) | (0.018) | (0.010) |
| *Salaried* | 0.022*** | 0.035*** | 0.037*** | 0.035*** | 0.009 | 0.030*** |
| | (0.005) | (0.005) | (0.006) | (0.005) | (0.014) | (0.010) |
| *Religion (Base: Hindus)* | | | | | | |
| *Muslims* | 0.002 | 0.003 | -0.030*** | 0.003 | -0.005 | 0.008 |
| | (0.006) | (0.005) | (0.005) | (0.005) | (0.014) | (0.009) |
| *Others* | 0.031*** | 0.091*** | 0.022* | 0.091*** | 0.035 | 0.092** |
| | (0.010) | (0.018) | (0.013) | (0.018) | (0.042) | (0.039) |
| *Urban* | -0.002 | 0.016*** | 0.013** | 0.016*** | -0.017 | 0.028*** |
| | (0.005) | (0.005) | (0.005) | (0.005) | (0.013) | (0.009) |
| *Constant* | -0.067*** | 0.115*** | -0.054*** | 0.115*** | 0.097 | 0.084 |
| | (0.010) | (0.019) | (0.019) | (0.019) | (0.086) | (0.052) |
| *State FE* | Y | Y | Y | Y | Y | Y |
| *1st stage F-stat* | | 207.282 | | 207.282 | | 207.282 |
| *Stock-Yogo critical value* | | 19.93 | | 19.93 | | 19.93 |
| *Sargan-Hansen* | | 0.266 | | 0.266 | | 0.266 |
| *J-statistic (p-val)* | | | | | | |
| *Observations* | 45,812 | 45,812 | 45,812 | 45,812 | 9,739 | 9,739 |
| *R-squared* | 0.048 | 0.319 | 0.090 | 0.319 | 0.099 | 0.394 |

Table 4: Effect of mobile financial services on financial outcomes-IV-Probit output

| | (1)<br>Investment-<br>2nd Stage | (2)<br>Investment-<br>1st Stage | (5)<br>Insurance-<br>2nd Stage | (6)<br>Insurance-<br>1st Stage | (9)<br>Borrowing-<br>2nd Stage | (10)<br>Borrowing-<br>1st Stage |
|---|---|---|---|---|---|---|
| *MFS* | 0.431*** | | 0.372*** | | 0.351*** | |
| | (0.077) | | (0.066) | | (0.128) | |
| *Tech ability* | | 0.388*** | | 0.387*** | | 0.440*** |
| | | (0.004) | | (0.004) | | (0.008) |
| *Share high education* | | 0.086*** | | 0.086*** | | 0.089*** |
| | | (0.011) | | (0.011) | | (0.022) |
| Controls | Y | Y | Y | Y | Y | Y |
| State FE | Y | Y | Y | Y | Y | Y |
| Observations | 45,714 | 45,714 | 45,812 | 45,812 | 9,706 | 9,706 |

Appendix

Table A1: Gender-wise effect of mobile financial services on financial outcomes –Second stage IV Probit output

| | (1) Investment-female | (2) Insurance-female | (3) Borrowing-female | (4) Investment-male | (5) Insurance-male | (6) Borrowing-male |
|---|---|---|---|---|---|---|
| MFS | 0.473*** | 0.345*** | 0.516*** | 0.389*** | 0.423*** | -0.019 |
| | (0.090) | (0.079) | (0.158) | (0.149) | (0.127) | (0.238) |
| Controls | Y | Y | Y | Y | Y | Y |
| State FE | Y | Y | Y | Y | Y | Y |
| Observations | 21,683 | 21,791 | 4,822 | 23,767 | 24,021 | 4,853 |

Table A2: Age-wise effect of mobile financial services on financial outcomes – Second stage IV Probit output

| | (1) Investment-young | (2) Insurance-young | (3) Borrowing-young | (4) Investment-old | (5) Insurance-old | (6) Borrowing-old |
|---|---|---|---|---|---|---|
| MFS | 0.405*** | 0.367*** | 0.406*** | 0.665*** | 0.467*** | 0.413 |
| | (0.088) | (0.075) | (0.148) | (0.172) | (0.161) | (0.307) |
| Controls | Y | Y | Y | Y | Y | Y |
| State FE | Y | Y | Y | Y | Y | Y |
| Observations | 31,195 | 31,266 | 6,260 | 14,385 | 14,519 | 3,429 |